\documentclass[aps,prd,preprint,tightenlines,superscriptaddress,showpacs]{revtex4}
\usepackage{graphicx}
\usepackage{epsfig}
\usepackage{dcolumn}
\usepackage{bm}
\usepackage{color}

\newcommand{\y}{Y(4260)}
\newcommand{\yy}{Y(4008)}
\newcommand{\yone}{Y(4360)}
\newcommand{\ytwo}{Y(4660)}
\newcommand{\lum}{{\cal L}}
\newcommand{\eff}{\varepsilon}
\newcommand{\BR}{{\cal B}}

\newcommand{\psp}{\psi(2S)}

\newcommand{\jpsi}{J/\psi}
\newcommand{\psift}{\psi(4040)}
\newcommand{\psifto}{\psi(4160)}
\newcommand{\psiftf}{\psi(4415)}
\newcommand{\EE}{e^+e^-}

\newcommand{\kk}{K^+K^-}

\newcommand{\ppjpsi}{\pi^+\pi^- J/\psi}

\newcommand{\pppsp}{\pi^+\pi^- \psp}

\newcommand{\beq}{\begin{equation}}
\newcommand{\eeq}{\end{equation}}
\newcommand{\bitm}{\begin{itemize}}
\newcommand{\eitm}{\end{itemize}}
\begin{document}

%************************************************************
\preprint{}
\title{Combined fit to BaBar and Belle data on $\EE\to \pppsp$}

  \affiliation{Institute of High Energy Physics, Chinese Academy of Sciences, Beijing 100049} % IHEP
  \affiliation{Shandong University, Jinan 250100}
  \affiliation{Huazhong Normal University, Wuhan 430079}

  \author{Z.~Q.~Liu}
  \affiliation{Institute of High Energy Physics, Chinese Academy of Sciences, Beijing 100049} % IHEP
  \affiliation{Huazhong Normal University, Wuhan 430079}
  \author{X.~S.~Qin}
  \affiliation{Institute of High Energy Physics, Chinese Academy of Sciences, Beijing 100049} % IHEP
  \affiliation{Shandong University, Jinan 250100}
  \author{C.~Z.~Yuan}
  \email{yuancz@ihep.ac.cn}
  \affiliation{Institute of High Energy Physics, Chinese Academy of Sciences, Beijing 100049} % IHEP

\date{\today}

\begin{abstract}

A combined fit is performed to the BaBar and Belle measurements of
the $\EE\to \pppsp$ cross sections for center-of-mass energy
between threshold and 5.5~GeV. The resonant parameters of the
$\yone$ and $\ytwo$ are determined. The mass is
$4355^{+9}_{-10}\pm 9$~MeV/$c^2$ and the width is
$103^{+17}_{-15}\pm 11$~MeV/$c^2$ for the $\yone$, and the mass is
$4661^{+9}_{-8}\pm 6$~MeV/$c^2$ and the width is
$42^{+17}_{-12}\pm 6$~MeV/$c^2$ for the $\ytwo$. The production of
the $\y$ in $\pppsp$ mode is found to be at $2\sigma$ level, and
$\BR(\y\to \pppsp)\cdot \Gamma_{\EE}$ is found to be less than
4.3~eV/$c^2$ at the 90\% confidence level, or equal to
$7.4^{+2.1}_{-1.7}$~eV/$c^2$ depending on it interferes with the
$\yone$ constructively or destructively. These information will
shed light on the understanding of the nature of the $Y$ states
observed in initial state radiation processes.

\end{abstract}


\maketitle

\section{Introduction}

Charmonium spectroscopy is of great interest to both the
experimentalists and the theorists since its first discovery more
than 30 years ago. With the successful running of the two
$B$-factories at SLAC and KEK, charmonium physics was also
revitalized as more and more charmonium and charmonium-like states
were observed in $B$ decays, in initial state radiation ($ISR$)
processes, in double-charmonium productions, and in two-photon
processes. The observation of the $X(3872)$~\cite{belle_x3872},
$\y$~\cite{babar_y4260,belle_y4260},
$Z(4430)^+$~\cite{belle_z4430} and so on may suggest the existence
of new type of hadronic states besides the conventional mesons
($q\bar{q}$) and baryons ($qqq$) in quark model. Among these many
newly observed states, those with $J^{PC}=1^{--}$ are becoming
very puzzling since there are too many of them between
4-5~GeV/$c^2$ than expected from the potential
models~\cite{eichten,godfrey}. In addition to the excited $\psi$
states observed in the inclusive hadronic cross
section~\cite{PDG,besres} (the $\psift$, $\psifto$, $\psiftf$),
there are four new structures observed in $ISR$ processes (the
$\yy$~\cite{belle_y4260}, $\y$~\cite{babar_y4260,belle_y4260},
$\yone$~\cite{babar,belle}, $\ytwo$~\cite{belle}). The
overpopulation of the vector states in this mass range may suggest
at least one of them are non-conventional charmonium
state~\cite{swanson}.

In the measurement of the $\pppsp$ cross section via $ISR$ at
BaBar~\cite{babar}, a broad resonance like structure was observed
with mass $4324\pm 24$~MeV/$c^2$ and width $172\pm 33$~MeV/$c^2$;
while with more luminosity, the Belle experiment found it is
indeed due to two narrow resonances with masses of 4361~MeV/$c^2$
and 4664~MeV/$c^2$~\cite{belle}. A close examination of the BaBar
observation shows that there is one bin with high cross section at
around the second resonance observed in Belle experiment although
it is not very significant.

In this paper, we try to perform a combined fit to $\EE\to \pppsp$
cross sections measured by the BaBar and Belle experiments with
two resonances, to obtain a better estimation of the resonant
parameters of the $\yone$ and $\ytwo$. We also study the possible
production of the $\y$ in $\pppsp$ mode.

\section{The data}

Both BaBar and Belle experiments reported cross sections of
$\EE\to \pppsp$ for center-of-mass energy ranges from threshold to
5.5~GeV. The integrated luminosity of the BaBar data sample is
$298~{\rm fb}^{-1}$ while that of the Belle data sample is
$673~{\rm fb}^{-1}$, with $\sim 90\%$ of the data were collected
at the $\Upsilon(4S)$ resonance ($\sqrt{s}=10.58$~GeV), while the
rest were taken off the $\Upsilon(4S)$ peak.

\begin{figure}
\includegraphics[width=9cm]{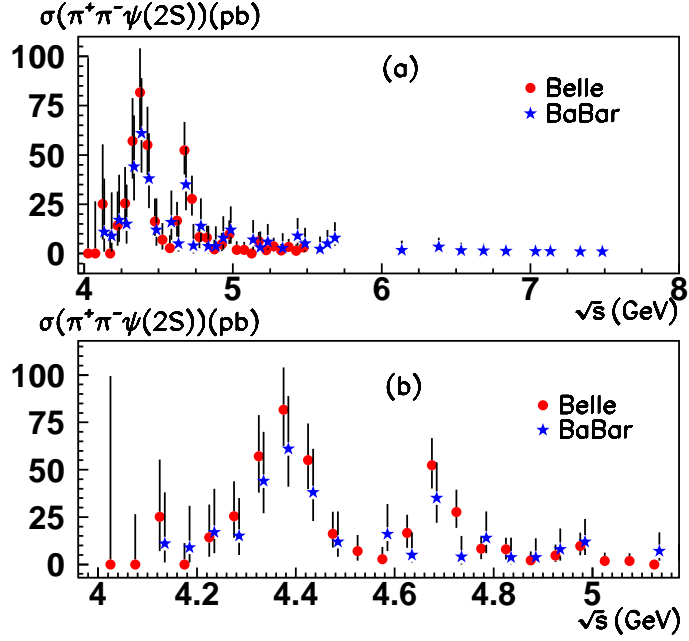}
\caption{\label{fig1} The cross sections of $\EE\to \pppsp$
measured at BaBar (stars with error bars) and Belle (dots with
error bars). (a) shows all the data, while (b) shows the region
with the $Y$ resonances.}
\end{figure}

Figure~\ref{fig1} shows the data, good agreement between BaBar and
Belle results is observed, and the two structures are evident.

\section{The formulae and the likelihood fit}

The Breit-Wigner form of a single resonance used in this analysis
is the same as in Refs.~\cite{babar,belle}, i.e.,
\begin{eqnarray}
\label{eq1}
BW(\sqrt{s})=\sqrt{\frac{M^{2}}{s}}\frac{\sqrt{12\pi\Gamma_{e^{+}e^{-}}
\BR(R\rightarrow f)\Gamma_{\rm tot}}}{s-M^{2}+iM\Gamma_{\rm tot}}
\sqrt{\frac{PS(\sqrt{s})}{PS(M)}} ,
\end{eqnarray}
where $M$ is the mass of the resonance, $\Gamma_{\rm tot}$ and
$\Gamma_{e^{+}e^{-}}$ are the total width and partial width to
$e^{+}e^{-}$ respectively, $\BR(R\rightarrow f)$ is the branching
fraction of $R$ decays into final state $f$, and $PS(\sqrt{s})$ is
the three-body decay phase space factor.

In fitting to the data, we assume all the cross sections are due
to the $\yone$ and $\ytwo$ resonances, and they are added
coherently, that is,
\begin{eqnarray}
\label{eq2}
\sigma(\sqrt{s})=|BW_{1}(\sqrt{s})+BW_{2}(\sqrt{s})\cdot
e^{i\phi}|^{2},
\end{eqnarray}
where $BW_{1}$ and $BW_{2}$ represent the two resonances and
$\phi$ is the relative phase between them.

We fit the data using a binned maximum likelihood method with
MINUIT in the CERN Program Library~\cite{minuit}. To take the
Poisson distribution of the small number of events in each
$\pppsp$ mass bin into consideration, we start from the observed
number of events in each mass bin instead of from the measured
cross section. For each $\pppsp$ mass bin, the probability of
observing $n^{\rm obs}_i$ events when the expected number of
events is $\lambda_i$ is $f_{i} = \frac{\lambda_{i}^{n_{i}^{\rm
obs}}e^{-\lambda_{i}}} {n_{i}^{\rm obs}!}$. Here
\begin{eqnarray}
\lambda_{i}=\sigma_{i}\eff_{i}\lum_{i}
\BR(\psi(2S)\rightarrow\pi^{+}\pi^{-}J/\psi)
\BR(J/\psi\rightarrow\ell^{+}\ell^{-})+n^{\rm bkg},
\end{eqnarray}
where $\sigma_{i}$ is the mean cross section in the $i$-th bin
(with bin width $\Delta m$), which is calculated as
\begin{eqnarray}
\sigma_{i}=\frac{1}{\Delta m}\int^{m+\Delta m/2}_{m-\Delta
m/2}\sigma(x)dx;
\end{eqnarray}
the number of background events in each bin, $n^{\rm bkg}=0.115$
for a 25~MeV/$c^2$ mass bin for Belle, and $n^{\rm bkg}$ is
neglected for BaBar. $\eff_{i}\lum_{i}
\BR(\psi(2S)\rightarrow\pi^{+}\pi^{-}J/\psi)
\BR(J/\psi\rightarrow\ell^{+}\ell^{-})$ for each bin is obtained
through $(n^{\rm obs}_{i}-n^{\rm bkg})/\sigma_{i}^{\rm exp}$ from
the $\pppsp$ invariant mass distributions and cross sections
presented in Refs.~\cite{babar,belle}.

The likelihood is defined as the product of $f_{i}$ over all the
$\pppsp$ mass bins, that is, \( {\cal L}=\prod_{i} f_{i} \). In
reality, $-2\ln{\cal L}$ is minimized to get the best estimation
of the parameters.

\section{The $\yone$ and $\ytwo$}

\begin{figure}
\includegraphics[width=9cm]{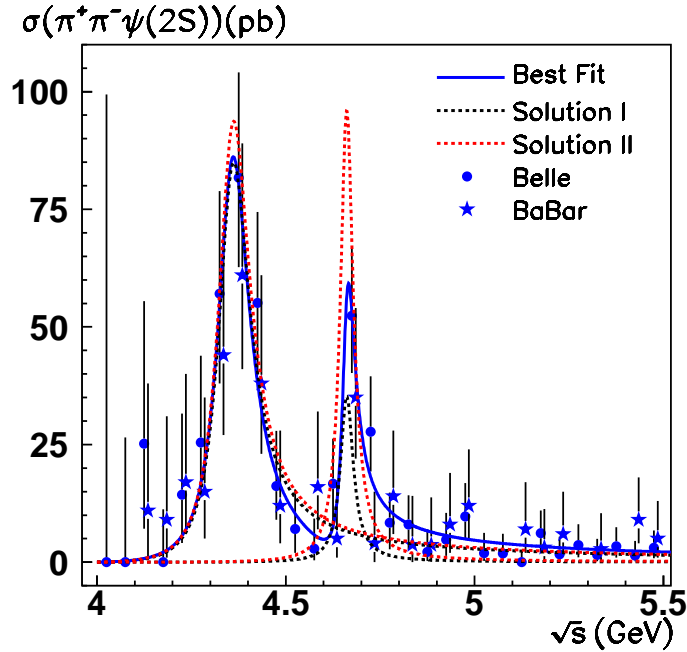}
\caption{\label{fig2} The results of the binned maximum likelihood
fit to $e^{+}e^{-}\rightarrow\pi^{+}\pi^{-}\psi(2S)$ data from
Belle and BaBar. The curves show the best fit with two coherent
Breit-Wigners and the contribution from each component. The
interference between the two amplitudes is not shown. The two
dashed curves at each peak show the two solutions (see text).}
\end{figure}

We fit the Belle data on $\pppsp$ between threshold and 5.5~GeV
(in 60 bins) and the BaBar data in the same energy range (in 30
bins) simultaneously. Figure~\ref{fig2} shows the fit results. Two
equally good solutions are found with the two amplitudes interfere
with each other differently. The masses and the widths of the two
resonances are identical but the partial widths to $e^{+}e^{-}$
and relative phases are different in these two solutions, as shown
in Table~\ref{table1}. The statistical significance of the $\ytwo$
is calculated by comparing the likelihood of the fit with and
without it, and we found it is $6.1\sigma$ in the combined fit,
while that quoted in Belle experiment is $5.8\sigma$~\cite{belle}.

\begin{table}
  \centering
\caption{\label{table1} Fit results to the combined BaBar and
Belle data on $e^{+}e^{-}\rightarrow\pi^{+}\pi^{-}\psi(2S)$. The
errors are statistical only. $M$, $\Gamma_{\rm tot}$, and
$\BR\Gamma_{e^{+}e^{-}}$ are the mass (in MeV/$c^{2}$), total
width (in MeV/$c^{2}$), and product of the branching fraction to
$\pi^{+}\pi^{-}\psi(2S)$ and the $e^{+}e^{-}$ partial width (in
eV/$c^{2}$), respectively. $\phi$ is the relative phase (in
degrees).}
\begin{tabular}{ccc}
  \hline\hline
  % after \\: \hline or \cline{col1-col2} \cline{col3-col4} ...
  Parameters & ~~~Solution I~~~ & ~~~Solution II~~~ \\
  \hline
  $M(\yone)$ &  \multicolumn{2}{c}{$4355^{+9}_{-10}$} \\
  $\Gamma_{\rm tot}(\yone)$ &  \multicolumn{2}{c}{$103^{+17}_{-15}$} \\
  $\BR\Gamma_{e^{+}e^{-}}(\yone)$ & $11.1^{+1.3}_{-1.2}$ & $12.3\pm 1.2$ \\
  $M(\ytwo)$ &  \multicolumn{2}{c}{$4661^{+9}_{-8}$}\\
  $\Gamma_{\rm tot}(\ytwo)$ & \multicolumn{2}{c}{$42^{+17}_{-12}$} \\
  $\BR\Gamma_{e^{+}e^{-}}(\ytwo)$ & $2.2^{+0.7}_{-0.6}$ & $5.9\pm 1.6$ \\
  $\phi$ & $18^{+23}_{-24}$ & $-74^{+16}_{-12}$ \\
  \hline\hline
\end{tabular}
\end{table}

To validate our fitting method, we fit the Belle data only with
two coherent Breit-Wigners, and the results are shown in
Table~\ref{table2} together with the Belle results~\cite{belle}
from an unbinned maximum likelihood fit method. From the table, we
found that while the two fits agree with each other reasonably
well, the mass of the second resonance shifts by about
9~MeV/$c^2$, which is comparable to the error quoted by Belle. One
of the reasons of the shift is the fitting method, that is, due to
the binning of the experimental data.

\begin{table}
 \centering
\caption{\label{table2} The binned fit to Belle data. The errors
are statistical only. The definitions of the parameters and the
units are the same as in Table~\ref{table1}. Numbers in
parentheses are taken from Ref.~\cite{belle} from an unbinned
fit.}
\begin{tabular}{ccc}
  \hline\hline
  % after \\: \hline or \cline{col1-col2} \cline{col3-col4} ...
  Parameters & ~~~~~~~~Solution I~~~~~~~~~~~~ & ~~~~~~~~~~~~Solution II~~~~~~~~ \\
  \hline
  $M(\yone)$ & \multicolumn{2}{c}{$4359^{+9}_{-10}$~($4361\pm 9$)} \\
  $\Gamma_{\rm tot}(\yone)$ &  \multicolumn{2}{c}{$85^{+18}_{-15}$~($74\pm 15$)} \\
  $\BR\Gamma_{e^{+}e^{-}}(\yone)$ & $11.9^{+2.5}_{-1.7}$~($10.4\pm 1.7$) & $12.8^{+1.7}_{-1.6}$~($11.8\pm 1.8$) \\
  $M(\ytwo)$ &  \multicolumn{2}{c}{$4655^{+11}_{-8}$~($4664\pm 11$)} \\
  $\Gamma_{\rm tot}(\ytwo)$ &  \multicolumn{2}{c}{$40^{+17}_{-15}$~($48\pm 15$)} \\
  $\BR\Gamma_{e^{+}e^{-}}(\ytwo)$ & $3.4^{+5.0}_{-0.9}$~($3.0\pm 0.9$) & $6.0^{+2.4}_{-3.5}$~($7.6\pm 1.8$) \\
  $\phi$ & $1^{+34}_{-79}$~($39\pm 30$) & $-56^{+91}_{-22}$~($-79\pm 17$) \\
  \hline\hline
\end{tabular}
\end{table}

\begin{figure}
\includegraphics[width=9cm]{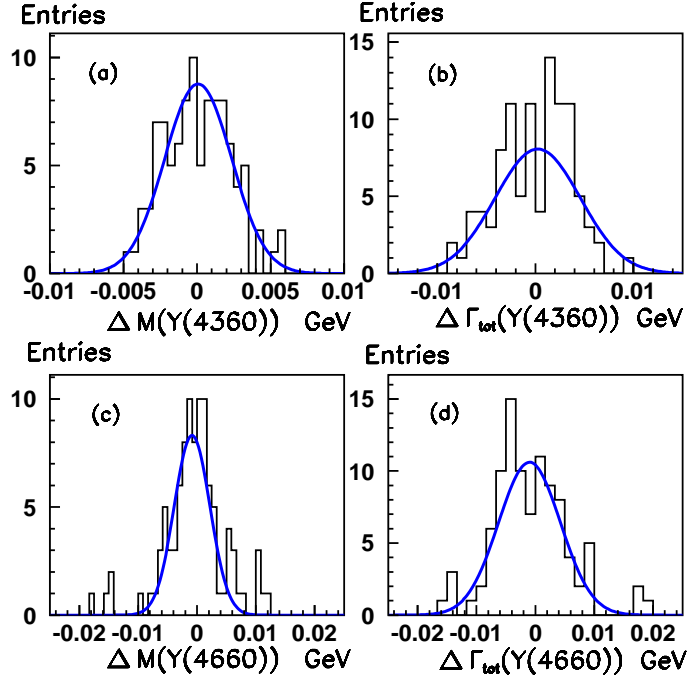}
\caption{\label{fig4} The distributions of the parameter
difference between binned and unbinned likelihood fits. The curves
show the fits with Gaussian functions. (a)~$M^{\rm
bin}(\yone)-M^{\rm unbin}(\yone)$, (b)~$\Gamma_{\rm tot}^{\rm
bin}(\yone)-\Gamma_{\rm tot}^{\rm unbin}(\yone)$, (c)~$M^{\rm
bin}(\ytwo)-M^{\rm unbin}(\ytwo)$, and (d)~$\Gamma_{\rm tot}^{\rm
bin}(\ytwo)-\Gamma_{\rm tot}^{\rm unbin}(\ytwo)$. }
\end{figure}

In principle, the difference between a binned and an unbinned
likelihood fit should be very small when the data sample is large
enough. However, in our special case, the data sample is rather
small (62 events in BaBar experiment and 110 events in Belle), and
thus the fluctuation due to binning is not small. In other words,
the lost of information in the binning process can not be
neglected. We test this by doing toy experiments with Monte Carlo
(MC) simulation. We generate 100 MC samples each with the same
number of observed events as in Belle experiment (110
events)~\cite{belle}. For each sample, we do binned and unbinned
likelihood fits and check the difference between the fit values of
the parameters. Figure~\ref{fig4} shows the distributions of the
differences in masses and widths from the two kinds of fit, and
Table~\ref{table3} shows the fit results to the distributions with
Gaussian functions.

It can be seen that the mean values of the differences are
consistent with zero as expected, while the standard deviations
are significantly different from zero, indicating that the
expected uncertainties introduced by using a binned fit are at a
few MeV/$c^2$ level from the unbinned fit. The observed difference
in the mass of the $\ytwo$ between binned and unbinned fits can be
explained as the uncertainty introduced by the binning procedure.
We take this kind of differences as one extra source of the
systematic error. It should be noted that the correlations between
the parameters are neglected and we do not try to quantize the
uncertainties of the $\BR\cdot \Gamma_{\EE}$ measurements here to
simplify the error estimation procedure.

\begin{table}
 \centering
\caption{\label{table3} Expected differences between binned and
unbinned fits. $\Delta M$ and $\Delta\Gamma_{\rm tot}$ are the
differences in mass (in MeV/$c^{2}$) and in total width (in
MeV/$c^{2}$). $\mu$ is the mean value and $\sigma$ is the standard
deviation of a Gaussian distribution, the errors are statistical.}
\begin{tabular}{ccc}
  \hline\hline
  Parameters & $\mu$ & $\sigma$ \\
  \hline
  $\Delta M(\yone)$ & $0.3\pm 0.3$ & $2.6\pm 0.3 $ \\
  $\Delta \Gamma_{\rm tot}(\yone)$ & $0.2\pm 0.6$ & $4.4\pm 0.5$ \\
  $\Delta M(\ytwo)$ & $-0.8\pm 0.4$ & $3.0\pm 0.5$\\
  $\Delta \Gamma_{\rm tot}(\ytwo)$ & $-1.0\pm 0.7$ & $5.2\pm 0.6$\\
  \hline\hline
\end{tabular}
\end{table}

The other sources of the systematic errors on the mass and width
measurements are not very different from those listed by the Belle
experiment~\cite{belle}, since it is mainly due to the
parametrization of the resonance. Taking directly the error from
Belle and adding in quadrature with the uncertainty due to the
binning, one gets the total systematic errors of 9~MeV/$c^2$,
11~MeV/$c^2$, 6~MeV/$c^2$, and 6~MeV/$c^2$, for $M(\yone)$,
$\Gamma_{\rm tot}(\yone)$, $M(\ytwo)$, and $\Gamma_{\rm
tot}(\ytwo)$, respectively.

\section{The $\y$}

The $\y$ was observed in $\ppjpsi$
mode~\cite{babar_y4260,belle_y4260}, and its production in
$\pppsp$ is not forbidden by any selection rule, except that the
phase space is a bit small due to large $\psp$ mass. We add the
$\y$ amplitude in Eq.~\ref{eq2} coherently with mass
(4247~MeV/$c^2$) and width (108~MeV/$c^2$) fixed to the Belle
measurement~\cite{belle_y4260} to measure its production rate.

\begin{figure}
\includegraphics[width=9cm]{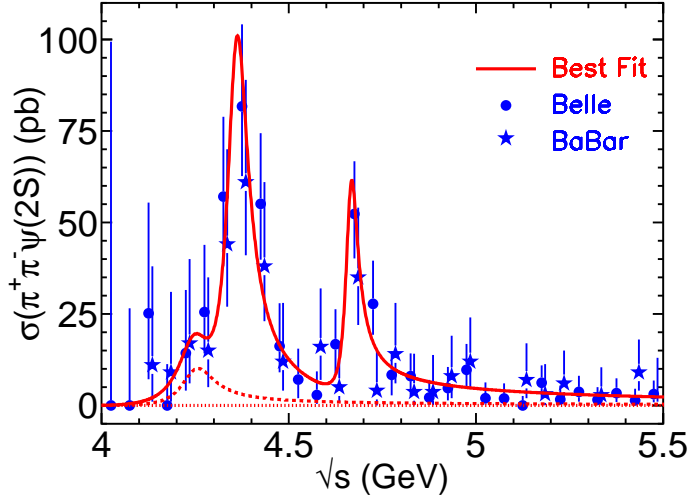}
\caption{\label{fig_y4260} The results of the fit to
$e^{+}e^{-}\rightarrow\pi^{+}\pi^{-}\psi(2S)$ data from Belle and
BaBar. The solid curve show the best fit with three coherent
Breit-Wigners: the $\y$, $\yone$, and $\ytwo$, and the dashed
curve is the signal shape of the $\y$.}
\end{figure}

Figure~\ref{fig_y4260} shows the fit result with three resonances,
the change of $-2\ln {\cal L}$ is 5.2 for two more free
parameters, this corresponds to a statistical significance of the
$\y$ of $1.8\sigma$. That is, the production of the $\y$ in
$\pppsp$ is not significant. There are two pairs of solutions for
$\BR(\y\to \pppsp)\cdot \Gamma_{\EE}$, one pair interferes with
the $\yone$ constructively, with $\BR(\y\to \pppsp)\cdot
\Gamma_{\EE} = 1.4^{+1.6}_{-0.9}$~eV/$c^2$, and we set the upper
limit as 4.3~eV/$c^2$ at the 90\% confidence level; the other pair
interferes with the $\yone$ destructively, and $\BR(\y\to
\pppsp)\cdot \Gamma_{\EE} = 7.4^{+2.1}_{-1.7}$~eV/$c^2$.

\section{Summary}

Based on combined Belle and BaBar data on $\EE\to \pppsp$, we
perform a maximum likelihood fit to get the resonant parameters of
the $\yone$ and $\ytwo$. We obtain $M(\yone)=4355^{+9}_{-10}\pm
9~{\rm MeV}/c^{2}$, $\Gamma_{\rm tot}(\yone)= 103^{+17}_{-15}\pm
11~{\rm MeV}/c^{2}$, $M(\ytwo)=4661^{+9}_{-8}\pm 6~{\rm
MeV}/c^{2}$, and $\Gamma_{\rm tot}(\ytwo)=42^{+17}_{-12}\pm 6~{\rm
MeV}/c^{2}$. These results give the best measurement of the
$\yone$ and $\ytwo$.

There is a faint evidence of $\y\to \pppsp$, with $\BR(\y\to
\pppsp)\cdot \Gamma_{\EE} < 4.3$~eV/$c^2$ at the 90\% confidence
level, or $\BR(\y\to \pppsp)\cdot \Gamma_{\EE} =
7.4^{+2.1}_{-1.7}$~eV/$c^2$, depending on the interference between
the $\y$ and $\yone$. These numbers should be compared with the
couplings of the $\y$ to $\ppjpsi$ in Ref.~\cite{belle_y4260} and
to $\kk\jpsi$ in Ref.~\cite{belle_kkjpsi}.

\acknowledgments

We thank Prof. S.~Olsen for suggesting extend the study to the
$\y$. This work is supported in part by the 100 Talents Program of
CAS under Contract No.~U-25 and by National Natural Science
Foundation of China under Contract No. 10491303.

\end{document}